\documentclass[english,pre,a4paper,twocolumn,noshowpacs,amsmath,amssymb]{revtex4}
\usepackage{graphicx,color}

\newcommand{\dd}{\mathrm{d}}

\newcommand{\pd}[2]{\frac{\partial #1}{\partial #2}}

\newcommand{\mean}[1]{\langle #1 \rangle}
\newcommand{\Int}[1]{\int\dd #1\;}
\newcommand{\IInt}[3]{\int_{#2}^{#3}\dd #1\;}

\newcommand{\z}{\omega}
\newcommand{\F}{\mathcal F}

\newcommand{\al}{\alpha}

\newcommand{\vhi}{\varphi}
\newcommand{\sig}{\sigma}

\newcommand{\ra}{\rightarrow}
\newcommand{\tri}{\triangleright}

\begin{document}

\title{Thermodynamic formalism
  and linear response theory \\ for non-equilibrium steady states}

\author{Thomas Speck}
\affiliation{Institut f\"ur Physik, Johannes Gutenberg-Universit\"at Mainz,
  Staudingerweg 7-9, 55128 Mainz, Germany}

\begin{abstract}
  We study the linear response in systems driven away from thermal equilibrium
  into a non-equilibrium steady state with non-vanishing entropy production
  rate. A simple derivation of a general response formula is presented under
  the condition that the generating function describes a transformation that
  (to lowest order) preserves normalization and thus describes a physical
  stochastic process. For Markov processes we explicitly construct the
  conjugate quantities and discuss their relation with known response
  formulas. Emphasis is put on the formal analogy with thermodynamic
  potentials and some consequences are discussed.
\end{abstract}

\maketitle


\section{Introduction}

One of the objectives of computational sciences is the accurate prediction of
material properties. The determination of transport coefficients (\emph{e.g.},
conductivities and mobilities) remains a challenge since, in general, it
implies currents and thus non-equilibrium conditions. Illustrative as well as
technologically important examples include the efficient transport of charges
in organic semiconductors~\cite{coro07,poel14} and across thin membranes in
reverse osmosis~\cite{kalr03}. While many sophisticated numerical methods have
been developed based on thermal equilibrium, for driven systems one typically
has to resort to brute-force computer experiments.

Sufficiently close to equilibrium transport coefficients can be determined
from equilibrium fluctuations via the fluctuation-dissipation
theorem~\cite{kubo}. There have been considerable efforts to find general
principles also for the linear response of non-equilibrium
states~\cite{spec06,baie09,chet09,pros09,spec10,seif11,warr12} (for more
complete reviews we refer to Refs.~\citenum{marc08,seif12,baie13} and
references therein), which find application in ``field-free'' numerical
algorithms~\cite{chat03,diez05,bert07}. There is now a ``zoo'' of different
approaches and derivations yielding (sometimes unrecognized) equivalent
results. One reason might be that actually several conjugate observables (and
their linear combinations) are equivalent in determining the
response~\cite{seif10}.

Extending the notion of statistical ensembles to trajectories (time-ordered
sequences of dynamic events) is currently receiving considerable
attention~\cite{leco07a,garr09,turn14}. A canonical structure for the joint
probability of \emph{microscopic} probabilities and their currents as been
formulated in Ref.~\citenum{maes08}. In contrast, here we are concerned with
macroscopic currents without information about microscopic probabilities (or
densities). Another concept is that of ``canonical'' path ensembles (also
appearing under the names $s$-ensemble~\cite{mero05,garr10}, tilted ensemble,
or Esscher transform) in which trajectories are biased by a time-integrated
observable. Under certain conditions typical trajectories in the canonical
path ensemble are equivalent to trajectories in the original processes with
fixed value of the observable~\cite{chet13,chet14,szav15,garr16}. The purpose
of this paper is to follow these ideas and apply them to the linear response
around a non-equilibrium steady state (NESS). It is organized as follows:
First, we briefly outline the canonical structure of intensive affinities and
extensive generalized distances for NESS. We then derive a general response
formula and show that it contains previously derived results, in particular
the response formula by Warren and Allen~\cite{warr12} and the path weight
representation~\cite{baie09,seif10,baie13}. Before concluding we discuss our
results in the light of a possible thermodynamic formalism for NESS.

\begin{figure}[b!]
  \centering
  \includegraphics{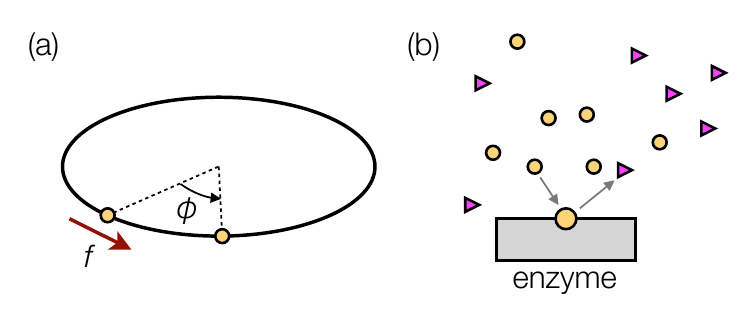}
  \caption{Examples for affinities and distances. (a)~Colloidal particle in a
    ring trap with radius $R$. The particle is driven by a constant force $f$
    (the affinity) while $X_\tau=R\phi$ is the total distance travelled during
    time $\tau$. (b)~Sketch of an enzyme driving the reaction
    $\circ\to\tri$. The generalized ``distance''
    $X_\tau=N^\tri_\tau=-N^\circ_\tau$ now corresponds to the number of $\tri$
    molecules produced during time $\tau$. The affinity
    $f=-(\mu^\tri-\mu^\circ)$ is given by the difference of chemical
    potential, which we assume to be fixed by chemiostats.}
  \label{fig:ness}
\end{figure}

\section{Thermodynamic formalism}

\subsection{Conjugate variables}

The mathematical structure of equilibrium statistical mechanics is based on
pairs of an extensive quantity (volume, particle number) and the conjugate
intensive quantity (pressure, chemical potential), which are related through
thermodynamic potentials (free energies). What makes the formalism so powerful
is that these potentials are also \emph{generating functions} encoding the
full statistics of the non-conserved extensive quantities. As a corollary,
fluctuations encode the response of thermodynamic observables to a small
external perturbation.

Pairs of \emph{apparently} conjugate quantities $(f^i,X^i)$ also arise for
non-equilibrium steady states (NESS), where non-zero intensive affinities
$f^i$ (the generalized forces) give rise to transport and thus extensive
(generalized) distances $X^i_\tau\sim\tau$ (measured over time $\tau$), see
Fig.~\ref{fig:ness} for two examples. Their product determines the entropy
production $\Sigma_\tau=\sum_if^iX^i_\tau$. Truly conjugate quantities,
however, would require the existence of a non-equilibrium ``potential''
$\Phi(f;\tau)$ so that $\mean{X^i_\tau}=\pd{\Phi}{f^i}$, which more generally
would determine state functions and justify variation
principles~\cite{mart06,sasa06}. Since this also implies strict convexity, it
would preclude established phenomena like a negative differential
mobility~\cite{jack08}.

\subsection{Linear response regime}
\label{sec:lr}

A thermodynamic description does, however, apply to the linear response
regime. To this end, consider a generalized distance $X_\tau\sim\tau$ measured
in thermal equilibrium (\emph{i.e.}, $f=0$) with probability distribution
$P_0(X;\tau)$. Clearly, the average $\mean{X_\tau}=0$ vanishes. The time
$\tau$ now plays a role similar to system size $N$ in conventional statistical
mechanics. We define the generating function
\begin{equation}
  \label{eq:Z}
  Z_0(f;\tau) \equiv \Int{X} e^{\frac{1}{2}fX} P_0(X;\tau)
  \asymp e^{\tau\frac{1}{2}\sig(f)}
\end{equation}
with large deviation function $\sig(f)$, where $\asymp$ denotes the asymptotic
limit of $\tau$ becoming larger than the longest correlation time. Following
the analogy with conventional thermodynamics we ask: Does $Z_0(f;\tau)$ for
$f\neq0$ describe the same physical system but now with non-zero affinity
(\emph{i.e.}, driven into a NESS)? A positive answer would imply that
\begin{equation}
  \mean{X_\tau} = \tau\pd{\sig}{f}
\end{equation}
holds, which, however, is not the case for arbitrary $f$. Only for small
$|f|\ll 1$ in the linear response regime does such an interpretation yield the
correct result with mean
\begin{equation}
  \label{eq:lr}
  \begin{split}
    \mean{X_\tau} &= \frac{1}{Z_0}\Int{X} Xe^{\frac{1}{2}fX}
    P_0(X;\tau) \\
    &= \frac{1}{2}\mean{(X_\tau)^2} f + \mathcal O(f^2).
  \end{split}
\end{equation}
This result is the well-known fluctuation-dissipation theorem~\cite{kubo}
through which fluctuations in thermal equilibrium determine how the system
reacts to a small applied force $f$. Indeed, Onsager's seminal insight has
been that in the linear response regime (half) ``the rate of increase of the
entropy plays the role of a potential''~\cite{onsa31}, namely the large
deviation function
\begin{equation}
  \label{eq:onsa}
  \sig(f) = \frac{1}{2}\sum_{ij}L^{ij}f^if^j
\end{equation}
with symmetric Onsager coefficients $L^{ij}=L^{ji}$ following from the
Green-Kubo relations
\begin{equation}
  \label{eq:L}
  \pd{}{\tau}\frac{1}{2}\mean{X^i_\tau X^j_\tau} \asymp L^{ij}.
\end{equation}

\subsection{Canonical path ensembles}

\begin{figure}[b!]
  \centering
  \includegraphics{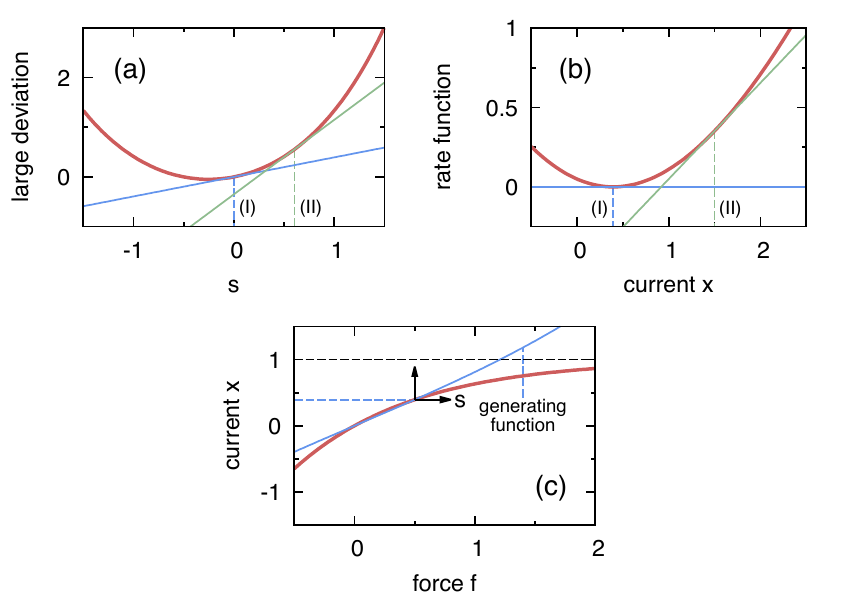}
  \caption{Illustration of the duality of $s$ and current $x=X_\tau/\tau$ for
    the asymmetric random walk (for details see Appendix~\ref{sec:arw}, with
    $f=\tfrac{1}{2}$ and $k^+=1$): (a)~Large deviation function
    $\phi_f^\ast(s)$ (thick line). The slope at $s=0$ (steady state I)
    corresponds to the mean current $x_0$. (b)~Rate function $\phi_f(x)$ of
    the current (thick line) with minimum at $x_0$. Conditioning currents to a
    value $x_s>x_0$ leads to the steady state (II) with $s>0$ determined by
    the slope. (c)~The unbiased average current as a function of driving force
    $f$ (thick line) and the current $x_s=\partial_s\phi_f^\ast$ (thin line)
    from the generating function, where $s=(f_\text{eff}-f)/2$ quantifies a
    perturbation of the steady state with effective force $f_\text{eff}$. Both
    currents agree for $s=0$ but deviate for increasing values of the
    perturbation $s$. For fixed $k^+$ the unbiased current is bounded (dashed
    black line).}
  \label{fig:arw}
\end{figure}

Away from the linear response regime for NESS characterized by the affinities
$f$ we can still define the generating function
\begin{equation}
  \label{eq:Z:f}
  Z_f(s;\tau) \equiv \Int{X} e^{sX} P_f(X;\tau) \asymp e^{\tau\phi_f^\ast(s)},
\end{equation}
where $s$ at this point is just the argument of this function. Moments and
cumulants are obtained through differentiation with respect to $s$ around
$s=0$. The function $\phi_f^\ast(s)$ is the large deviation function, which by
construction is a convex function. It is related to the rate function
$P_f(X;\tau)\asymp e^{-\tau\phi_f(x)}$ for the current $x\equiv X_\tau/\tau$ through
the Legendre-Fenchel transform~\cite{touc09}
\begin{equation}
  \phi_f^\ast(s) = \sup_x[xs-\phi_f(x)].
\end{equation}
One can now ask the following question: Assume that we condition the path
ensemble to contain only trajectories with a fixed value $x_s$ for the
current. As discussed in detail by Chetrite and
Touchette~\cite{chet13,chet14}, in the limit of large $\tau$ this
``microcanonical'' ensemble becomes equivalent (under mild assumptions) to a
``canonical'' ensemble in which the current fluctuates but its mean equals
$x_s$. This canonical path ensemble is described through the generating
function Eq.~(\ref{eq:Z:f}) for a value of $s$ determined through the
condition $s=\partial_x\phi_f|_{x_s}$. While $s$ and $x$ are thus conjugate
quantities (see Fig.~\ref{fig:arw} for an illustration for a specific system),
changing $s$ does not trace a change of the affinities but involves a rather
complicated, non-local transformation (Doob's transform) also of the
interactions~\cite{jack10a,spec11,chet13,chet14,nyaw16}. It is exactly this
behavior that complicates a general thermodynamic description of
NESS. However, in the following we demonstrate that small $s$ can be
interpreted as a perturbation of the steady state, leading trivially to linear
response relations. Moreover, based on this result we can construct a
different set of conjugate quantities which extend Onsager's result to NESS
driven beyond the linear response regime.


\section{A general response formula}

We consider a steady state maintained by at least one non-vanishing affinity
and thus having a non-vanishing entropy production rate
$\mean{\Sigma_\tau}>0$. For clarity, in this section we consider a single
perturbation but the generalization to more than one is straightforward. To be
sufficiently general, we define our quantity of interest through the
stochastic (Riemann-Stieltjes) integral of the form
\begin{equation}
  \label{eq:R}
  R_\tau[\z_t] \equiv \IInt{r(\z_t)}{0}{\tau} 
  = \IInt{t}{0}{\tau} \dot r(\z_t,\dot\z_t)
\end{equation}
over a process $\z_t$ representing the state $\z$ of the system at time
$t$. The integral maps a single trajectory $\{\z_t\}_0^\tau$ of length
$\tau\geqslant 0$ onto a real number. Clearly, $R_\tau\sim\tau$ is a
time-extensive quantity. It will be convenient later to also introduce the
generalized velocity $\dot r(\z,\dot\z)$, which for stochastic processes is to
be understood symbolically and follows the notational convention that
typically is used in physics.

The joint probability $\phi(\z,R,t)$ describes the probability to observe the
system in state $\z$ at time $t$ having accumulated an amount $R=R_t$ up to
time $t\leqslant\tau$ starting with $R_0=0$. Hence, the initial condition
factorizes to $\phi(\z,R,0)=\psi(\z)\delta(R)$, where $\psi(\z)$ is the steady
state probability of state $\z$ and $\delta(R)$ is the Dirac
$\delta$-function.

It is often more convenient to work with the (Laplace) transform
\begin{equation}
  \label{eq:laplace}
  \hat\phi_s(\z,t) \equiv \Int{R} e^{sR}\phi(\z,R,t)
\end{equation}
with initial condition $\hat\phi_s(\z,0)=\psi(\z)$ following from the
factorization of the joint probability, where the integral runs over all
possible values of $R$. Moreover, for $s=0$ we have that
$\hat\phi_0(\z,t)=\psi(\z)$ is the steady state probability. While in general
$\hat\phi_s(\z,t)$ is not normalized, we now explore the consequences of
\emph{demanding} that $\hat\phi_s(\z,t)$ remains a normalized function for
$s\neq0$. Since for non-negative $\phi(\z,R,t)$ Eq.~(\ref{eq:laplace}) implies
that also $\hat\phi_s(\z,t)$ is non-negative, it can be interpreted as the
probability distribution of state $\z$ for a system parametrized by $s$. At
this point the physical meaning of $s$ is not obvious but in the next section
we will construct explicitly conjugate pairs $(s,R)$.

Now consider a system described by the probability distribution
$\hat\phi_s(\z,t)$. The expectation value for an arbitrary observable $A(\z)$
becomes
\begin{equation}
  \mean{A(t)}_s = \sum_\z A(\z)\hat\phi_s(\z,t),
\end{equation}
which reduces to $\mean{A}$ for $\hat\phi_0(\z,t)=\psi(\z)$ at $s=0$. Hence,
in the following we interpret the conjugate variable $s$ appearing in the
generating function to describe an external perturbation applied to the system
at $t=0$ and driving it towards a neighboring steady state. The response to
this perturbation is
\begin{equation}
  \label{eq:resp}
  \begin{split}
    \left.\pd{\mean{A(\tau)}_s}{s}\right|_{s=0} &= 
    \sum_\z A(\z)\left.\pd{\hat\phi_s(\z,\tau)}{s}\right|_{s=0} \\
    &= \sum_\z \Int{R} A(\z)R\phi(\z,R,\tau) \\
    &= \mean{A(\tau)R_\tau},    
  \end{split}
\end{equation}
which follows inserting Eq.~(\ref{eq:laplace}). This is our central result. It
relates the response (sometimes called sensitivity) of an observable to the
correlations of this observable with the amount of $R$ accumulated since the
perturbation was applied. The correlations are to be determined in the
unperturbed steady state corresponding to $s=0$. The result
Eq.~(\ref{eq:resp}) is quite general and does not require any assumptions on
the dynamics.


\section{Constructing conjugate pairs}

The response Eq.~(\ref{eq:resp}) follows for functions $\hat\phi_s(\z,t)$
that, at least for small $s$, are normalized. This places some restrictions
onto what integrals $R_\tau[\z_t]$ are actually admissible. One property
follows immediately by choosing $A(\z)=1$, which implies that the average
$\mean{R_t}=0$ vanishes for any $t>0$.

\subsection{Stochastic dynamics}

To be more specific, we consider a continuous Markov process
\begin{equation}
  \label{eq:lang}
  \dd\z_t = \F(\z_t)\dd t + \dd\xi(\z_t)
\end{equation}
with effective drift vector $\F(\z)$ and random increments
\begin{equation}
  \dd\xi(\z) = \sum_\al\sig_\al(\z)\circ\dd W_\al(t),
\end{equation}
where $W_\al(t)$ are independent Wiener processes and the symbol $\circ$
denotes the Stratonovich rule for stochastic integrals. With the symmetric
diffusion matrix
\begin{equation}
  \label{eq:D}
  D^{ij}(\z) \equiv \frac{1}{2}\sum_\al \sig_\al^i(\z)\sig_\al^j(\z)
\end{equation}
the Markov generator reads
\begin{equation}
  \label{eq:L:0}
  L_0 = F\cdot D\cdot\nabla + \nabla\cdot D\cdot\nabla = (F+\nabla)\cdot
  D\cdot\nabla.
\end{equation}
Its adjoint $L^\dagger_0$ generates the time evolution of the probability
distribution, $\partial_t\psi=L^\dagger_0\psi$. Here, $F(\z)$ is the physical
force such that the effective drift becomes
\begin{equation}
  \F = D\cdot F + \frac{1}{2}\sum_\al(\nabla\cdot\sig_\al)\sig_\al.
\end{equation}
Throughout we set Boltzmann's constant and temperature to unity so that
entropies are dimensionless and the mobility matrix coincides with the
diffusion matrix Eq.~(\ref{eq:D}).

\subsection{The response formula of Warren and Allen}

We first consider
\begin{equation}
  \dd r(\z) = h(\z)\dd t + g(\z)\circ\dd\xi(\z),
\end{equation}
where the vector $g(\z)$ couples to the same noise as in
Eq.~(\ref{eq:lang}). Following Chetrite and Touchette~\cite{chet13,chet14},
the \emph{tilted} (or deformed) generator for the evolution
$\partial_t\hat\phi_s=L^\dagger_s\hat\phi_s$ becomes
\begin{equation}
  \label{eq:L:s}
  L_s = F\cdot D\cdot\nabla + (\nabla+sg)\cdot D\cdot(\nabla+sg)+sh,
\end{equation}
which for $s=0$ reduces to the generator $L_0$ in
Eq.~(\ref{eq:L:0}). Expanding the second term to linear order of $s$, we can
recast this generator into the form
\begin{equation}
  \label{eq:L:s:lin}
  L_s = (F+2sg+\nabla)\cdot D\cdot\nabla + s[\nabla\cdot(D\cdot g)+h]
  + \mathcal O(s^2),
\end{equation}
which manifestly preserves normalization if $\nabla\cdot(D\cdot g)+h=0$, which
thus determines $h(\z)$. Note that changing from Stratonovich to It\^{o}
calculus, this condition implies that $\dd r=g\cdot\dd\xi$. Clearly, since
then state and noise are independent, the expectation value of $R$ vanishes as
required.

We now assume that the perturbed steady state is described by the forces
$F_s(\z)$ depending on $s$. Expanding the forces to linear order,
\begin{equation}
  \label{eq:F}
  F_s(\z) = F(\z) + s\left.\pd{F_s(\z)}{s}\right|_{s=0} + \mathcal O(s^2),
\end{equation}
we read off the coefficient
\begin{equation}
  \label{eq:g}
  g(\z) = \frac{1}{2}\left.\pd{F_s(\z)}{s}\right|_{s=0}
\end{equation}
by comparing with Eq.~(\ref{eq:L:s:lin}). This is the result found in
Ref.~\citenum{warr12} following quite a different approach. Provided we know
how the forces depend on $s$, we have thus constructed one possible observable
$R_\tau$ to be used in Eq.~(\ref{eq:resp}).

\subsection{Coupling to state changes}

A more general form of time-extensive observables is given by
\begin{equation}
  \label{eq:r}
  \dd r(\z) = h(\z)\dd t + g(\z)\circ\dd\z,
\end{equation}
where the vector $g(\z)$ now couples to the evolution of the state $\z$. The
generator follows as
\begin{equation}
  L_s = F\cdot D\cdot(\nabla+sg) + (\nabla+sg)\cdot D\cdot(\nabla+sg)+sh.
\end{equation}
Expanding to lowest order we again find Eq.~(\ref{eq:g}) for the coefficient
$g(\z)$ and the condition to preserve normalization now becomes
\begin{equation}
  \label{eq:h}
  F\cdot D\cdot g + \nabla\cdot(D\cdot g) + h = 0.
\end{equation}
It is straightforward to check that the time-integrated observable $R_\tau$
following from Eq.~(\ref{eq:r}) can be written as the derivative
\begin{equation}
  R_\tau = -\left.\pd{\mathcal A_\tau}{s}\right|_{s=0}
\end{equation}
with stochastic action
\begin{equation}
  \mathcal A_\tau[\z_t] \equiv \IInt{t}{0}{\tau} \mathcal L_s(\z_t,\dot\z_t),
\end{equation}
where (still employing the Stratonovich rule)
\begin{multline}
  \mathcal L_s(\z,\dot\z) = \frac{1}{4}(\dot\z-D\cdot F_s)\cdot
  D^{-1}\cdot(\dot\z-D\cdot F_s) \\ + \frac{1}{2}\nabla\cdot(D\cdot F_s).
\end{multline}
Hence, employing Eq.~(\ref{eq:r}), the conjugate observable $R_\tau$ now
corresponds to the ``path weight representation'' discussed in
Refs.~\citenum{baie09,seif10,baie13}.


\section{Discussion}

\subsection{Thermodynamic formalism}

It is straightforward to extend Eq.~(\ref{eq:resp}) to multiple affinities
$s=\{s^i\}$. We restrict our considerations to the set of observables
$\{R^i_\tau\}$ with vanishing mean, for which we can derive a \emph{local}
potential. To this end, from the transformed joint probability
Eq.~(\ref{eq:laplace}) we define the generating function
\begin{equation}
  Z_f(s;\tau) = \sum_\z \hat\phi_s(\z,\tau) \asymp e^{\tau\vhi_f^\ast(s)},
\end{equation}
where we make explicit the dependency on the affinities $f$ driving the
system. In the limit $\tau\to\infty$ the large deviation function again
follows from the Legendre-Fenchel transform
\begin{equation}
  \vhi_f^\ast(s) = \sup_r \left[r\cdot s - \vhi_f(r) \right],
\end{equation}
where we have assumed that a large deviation principle
$P_f(R;\tau)\asymp e^{-\tau\vhi_f(r)}$ holds with $r^i\equiv R^i/\tau$. As a
consequence, $\vhi_f^\ast(s)$ is \emph{always} a convex function and
constitutes our local potential around a steady state determined by the
affinities $f$. For a potential $\vhi_f^\ast(s)$ that is differentiable at
$s=0$, the correlations are manifestly symmetric and follow as
\begin{equation}
  \mean{R^i_\tau R^j_\tau} = \left.\pd{^2}{s^i\partial s^j}\ln
    Z_f\right|_{s=0} \asymp
  \left.\tau\pd{^2\vhi_f^\ast}{s^i\partial s^j}\right|_{s=0} = \tau L_f^{ij}
\end{equation}
with steady state susceptibilities
\begin{equation}
  L_f^{ij} \equiv \lim_{\tau\ra\infty}
  \frac{1}{\tau}\pd{\mean{R^i_\tau}_s}{s^j} = \pd{\mean{\dot r^i}_s}{s^j}.
\end{equation}
This result extends the Onsager potential Eq.~(\ref{eq:onsa}) to non-zero
affinities and emphasizes the canonical structure. An interesting consequence
is that, employing Legendre transforms as in conventional thermodynamics, we
can now switch between affinity $s^i$ and current $\mean{\dot r^i}$ depending
on what is the more convenient variable for a specific situation. Moreover,
susceptibilities are related by Maxwell and further relations (similar to,
\emph{e.g.}, the relation between the heat capacities at constant volume and
constant pressure).

\subsection{Illustration: Single particle in a ring trap}

To briefly illustrate our results we consider the paradigmatic single
colloidal particle moving in a ring trap~\cite{blic07,gome09,gome11,nyaw16},
see Fig.~\ref{fig:ness}(a). The state of the system is given by the position
$x$ with force $F(x)=-\partial_xU(x)+f$, where $U(x)$ is an external, periodic
potential energy and $f$ is the constant driving force. The diffusion
coefficient $D_0$ is independent of $x$. The particle is driven into the
unperturbed NESS through the force $f$ with non-zero average speed
$v(f)=\mean{\dot x}=D_0[f-\mean{\partial_xU}]$. As perturbation we consider a
change of the driving force, $f\to f+s$, with
\begin{equation}
  F_s(x) = -\partial_xU(x) + f + s.
\end{equation}
From Eq.~(\ref{eq:F}) we immediately find 
\begin{equation}
  g = \left.\frac{1}{2}\pd{F_s}{s}\right|_{s=0} = \frac{1}{2}.
\end{equation}
From Eq.~(\ref{eq:h}) one then obtains $h(x)=-\frac{1}{2}D_0F_f(x)$ and thus
from Eq.~(\ref{eq:r}) the generalized velocity
\begin{equation}
  \label{eq:trap:r}
  \dot r(x,\dot x) = \frac{1}{2}[\dot x-D_0F_f(x)].
\end{equation}
This is indeed one of the admissible choices for determining the response with
respect to a change of the driving force~\cite{seif10}.

The average of Eq.~(\ref{eq:trap:r}) for a perturbed NESS with $f+s$ becomes
\begin{equation}
  \mean{\dot r}_s = \frac{1}{2}[v(f+s)+D_0\mean{\partial_xU}_s-D_0f] =
  \frac{1}{2} D_0s
\end{equation}
after inserting the speed
$v(f+s)=\mean{\dot x}_s=D_0[-\mean{\partial_xU}_s+f+s]$. Due to the additivity
of the perturbation, the conjugate variable $\mean{\dot r}$ is a simple linear
function of $s$ independent of $f$ implying the potentials
$\vhi^\ast(s)=\frac{1}{4}D_0s^2$ and $\vhi(r)=\frac{1}{D_0}r^2$. Hence, while
$v(f)$ is a non-linear function of the driving force $f$, the local potential
describes trivial, equilibrium-like fluctuations~\cite{chet09}. Close to
equilibrium in the linear response regime one recovers
$\mean{\dot x}=2\mean{\dot r}=D_0f$ as expected.

\subsection{Fluctuation theorem}

What is the physical meaning of $s$? To get some insight let us assume that
$s$ shifts the steady state to $f+s$ with probability
\begin{equation}
  \label{eq:prob}
  P_{f+s}(R;\tau) \sim e^{sR}P_f(R;\tau),
\end{equation}
which is the expression that appears in the generating function. For a large
class of observables (including the example from the previous subsection) we
can write $R_\tau=\frac{1}{2}(X_\tau-S_\tau)$ as a current $X_\tau$ minus
another term $S_\tau$, both of which have the same average
$\mean{X_\tau}=\mean{S_\tau}$ in the unperturbed NESS. While the current is
antisymmetric with respect to time reversal, $X^\dagger_\tau=-X_\tau$, the
second term $S^\dagger_\tau=S_\tau$ is invariant. The fluctuation
theorem~\cite{seif12} then becomes
\begin{equation}
  \frac{P_{f+s}(R;\tau)}{P_{f+s}(R^\dagger;\tau)} =
  \frac{P_f(X,S;\tau)}{P_f(-X,S;\tau)} e^{sX} = e^{(f+s)X},
\end{equation}
where the final expression involves the entropy $\Sigma_\tau=(f+s)X_\tau$
produced in the perturbed NESS. This shows that the parameter $s$ of the
generating function, for the pair $(s,R)$, indeed corresponds to a change of
the affinity $f$ determining the unperturbed NESS. The importance of the
time-symmetric contribution $S_\tau$ for the non-equilibrium linear response
has been discussed by C. Maes and coworkers~\cite{baie09,basu15}.

\subsection{Linear response regime}

As eluded to in the introduction, the observable $R_\tau$ appearing in
Eq.~(\ref{eq:resp}) is not unique. This becomes apparent in the linear
response regime perturbing thermal equilibrium when choosing the current
$R_\tau\to X_\tau$, which also has vanishing mean. Again appealing to the
fluctuation theorem we have for small $|f|\ll1$
\begin{equation}
  \frac{P_f(X;\tau)}{P_f(-X;\tau)} =
  \frac{e^{\frac{1}{2}f\cdot X}P_0(X;\tau)}{e^{-\frac{1}{2}f\cdot X}P_0(-X;\tau)}
  = e^{fX}.
\end{equation}
Here we have used that the currents change sign under time reversal, whereby
in equilibrium $P_0(-X;\tau)=P_0(X;\tau)$ holds. Following Eq.~(\ref{eq:prob})
one sees that now we have to use $s\to\frac{1}{2}f$ leading to the definition
of the generating function Eq.~(\ref{eq:Z:f}) given in Sec.~\ref{sec:lr},
which in turn leads to the famous Onsager result.


\section{Conclusions}

In this paper we have studied the tilted Markov generator under the condition
that for small tilt $s$ it preserves normalization and thus describes a
physical stochastic process. Identifying this process as a shifted steady
state has allowed us to interpret the abstract tilt parameter $s$ of the
generating function as a perturbation of the original steady state. For Markov
processes we have explicitly constructed two types of conjugate observables
$R$ that encode the system's response and thus allow to determine transport
coefficients from correlations in the unperturbed steady state. Only forces
are required as input, no explicit knowledge of the stationary distribution or
entropy production is necessary.

What is perhaps most interesting is the notion of different ensembles
analogous to conventional thermodynamics. Consider for example the situation
that we require a transversal transport coefficient for fixed longitudinal
field (affinity) although the simulations (or experiments) have to be
performed at fixed current. Transport coefficients in one ensemble could then
be calculated from those in another ensemble much in the same way the heat
capacity at constant pressure is calculated from the heat capacity at constant
volume. The approach presented here might pave the way for a systematic theory
although the simple example of a trapped Brownian particle demonstrates that
not all informations about the steady state are encoded in the corresponding
local potential.


\appendix

\section{Asymmetric random walk}
\label{sec:arw}

As a specific example we consider the asymmetric random walk
(ARW)~\cite{lebo99,mehl08,spec12}, for which we can perform the
transformations analytically. The ARW describes the motion of a walker on an
infinite lattice [cf. Fig.~\ref{fig:ness}(a)] with discrete sites. The walker
jumps forward and backward with rates $k^+$ and $k^-$, respectively. The
affinity is simply the force $f=\ln(k^+/k^-)$. For $N^+_\tau$ steps forward
and $N^-_\tau$ steps backward, the distance traveled is
$X_\tau=N^+_\tau-N^-_\tau$ with average
\begin{equation}
  \label{eq:arw}
  \mean{X_\tau} = \tau(k^+-k^-) = \tau k^+(1-e^{-f}).
\end{equation}
Note that here the distance takes only discrete integer values. Its
probability is known analytically~\cite{kampen}
\begin{equation}
  P_f(X;\tau) = I_X(2\sqrt{k^+k^-}\tau)(k^+/k^-)^{X/2}e^{-(k^++k^-)\tau},
\end{equation}
where $I_n(z)$ is the modified Bessel function of the first kind of order
$n$. The generating function
\begin{equation}
  Z_f(s;\tau) = \sum_{X=-\infty}^\infty e^{sX} P_f(X;\tau)
\end{equation}
can be calculated exactly using~\cite{abramowitz}
\begin{equation}
  \sum_{X=-\infty}^\infty I_X(z)c^X = \exp\left[(z/2)(c+c^{-1})\right].
\end{equation}
The result is $Z_f(s;\tau)=\exp[\tau\phi_f^\ast(s)]$ with
\begin{equation}
  \phi_f^\ast(s) = k^+\left(e^s+e^{-(f+s)}-e^{-f}-1\right).
\end{equation}
Clearly, the derivative
\begin{equation}
  x_s = \pd{\phi_f^\ast}{s} = k^+\left(e^s - e^{-(f+s)}\right)
\end{equation}
only agrees with the current Eq.~(\ref{eq:arw}) for $s=0$. Note that for this
simple example the same current can be achieved through the effective force
$f_\text{eff}=f+2s$ while simultaneously rescaling time. The slightly more
complex example of a bias random walker with two internal states has been
treated in Ref.~\citenum{spec11}.



\begin{thebibliography}{45}
\expandafter\ifx\csname natexlab\endcsname\relax\def\natexlab#1{#1}\fi
\expandafter\ifx\csname bibnamefont\endcsname\relax
  \def\bibnamefont#1{#1}\fi
\expandafter\ifx\csname bibfnamefont\endcsname\relax
  \def\bibfnamefont#1{#1}\fi
\expandafter\ifx\csname citenamefont\endcsname\relax
  \def\citenamefont#1{#1}\fi
\expandafter\ifx\csname url\endcsname\relax
  \def\url#1{\texttt{#1}}\fi
\expandafter\ifx\csname urlprefix\endcsname\relax\def\urlprefix{URL }\fi
\providecommand{\bibinfo}[2]{#2}
\providecommand{\eprint}[2][]{\url{#2}}

\bibitem[{\citenamefont{Coropceanu et~al.}(2007)\citenamefont{Coropceanu,
  Cornil, da~Silva~Filho, Olivier, Silbey, and Br\'edas}}]{coro07}
\bibinfo{author}{\bibfnamefont{V.}~\bibnamefont{Coropceanu}},
  \bibinfo{author}{\bibfnamefont{J.}~\bibnamefont{Cornil}},
  \bibinfo{author}{\bibfnamefont{D.~A.} \bibnamefont{da~Silva~Filho}},
  \bibinfo{author}{\bibfnamefont{Y.}~\bibnamefont{Olivier}},
  \bibinfo{author}{\bibfnamefont{R.}~\bibnamefont{Silbey}}, \bibnamefont{and}
  \bibinfo{author}{\bibfnamefont{J.-L.} \bibnamefont{Br\'edas}},
  \bibinfo{journal}{Chem. Rev.} \textbf{\bibinfo{volume}{107}},
  \bibinfo{pages}{926} (\bibinfo{year}{2007}).

\bibitem[{\citenamefont{Poelking et~al.}(2014)\citenamefont{Poelking, Tietze,
  Elschner, Olthof, Hertel, Baumeier, W\"urthner, Meerholz, Leo, and
  Andrienko}}]{poel14}
\bibinfo{author}{\bibfnamefont{C.}~\bibnamefont{Poelking}},
  \bibinfo{author}{\bibfnamefont{M.}~\bibnamefont{Tietze}},
  \bibinfo{author}{\bibfnamefont{C.}~\bibnamefont{Elschner}},
  \bibinfo{author}{\bibfnamefont{S.}~\bibnamefont{Olthof}},
  \bibinfo{author}{\bibfnamefont{D.}~\bibnamefont{Hertel}},
  \bibinfo{author}{\bibfnamefont{B.}~\bibnamefont{Baumeier}},
  \bibinfo{author}{\bibfnamefont{F.}~\bibnamefont{W\"urthner}},
  \bibinfo{author}{\bibfnamefont{K.}~\bibnamefont{Meerholz}},
  \bibinfo{author}{\bibfnamefont{K.}~\bibnamefont{Leo}}, \bibnamefont{and}
  \bibinfo{author}{\bibfnamefont{D.}~\bibnamefont{Andrienko}},
  \bibinfo{journal}{Nature Mater.} \textbf{\bibinfo{volume}{14}},
  \bibinfo{pages}{434–} (\bibinfo{year}{2014}).

\bibitem[{\citenamefont{Kalra et~al.}(2003)\citenamefont{Kalra, Garde, and
  Hummer}}]{kalr03}
\bibinfo{author}{\bibfnamefont{A.}~\bibnamefont{Kalra}},
  \bibinfo{author}{\bibfnamefont{S.}~\bibnamefont{Garde}}, \bibnamefont{and}
  \bibinfo{author}{\bibfnamefont{G.}~\bibnamefont{Hummer}},
  \bibinfo{journal}{Proc. Natl. Acad. Sci. U.S.A.}
  \textbf{\bibinfo{volume}{100}}, \bibinfo{pages}{10175}
  (\bibinfo{year}{2003}).

\bibitem[{\citenamefont{Kubo et~al.}(1991)\citenamefont{Kubo, Toda, and
  Hashitsume}}]{kubo}
\bibinfo{author}{\bibfnamefont{R.}~\bibnamefont{Kubo}},
  \bibinfo{author}{\bibfnamefont{M.}~\bibnamefont{Toda}}, \bibnamefont{and}
  \bibinfo{author}{\bibfnamefont{N.}~\bibnamefont{Hashitsume}},
  \emph{\bibinfo{title}{Statistical Physics II}}
  (\bibinfo{publisher}{Springer-Verlag}, \bibinfo{address}{Berlin},
  \bibinfo{year}{1991}), \bibinfo{edition}{2nd} ed.

\bibitem[{\citenamefont{Speck and Seifert}(2006)}]{spec06}
\bibinfo{author}{\bibfnamefont{T.}~\bibnamefont{Speck}} \bibnamefont{and}
  \bibinfo{author}{\bibfnamefont{U.}~\bibnamefont{Seifert}},
  \bibinfo{journal}{Europhys. Lett.} \textbf{\bibinfo{volume}{74}},
  \bibinfo{pages}{391} (\bibinfo{year}{2006}).

\bibitem[{\citenamefont{Baiesi et~al.}(2009)\citenamefont{Baiesi, Maes, and
  Wynants}}]{baie09}
\bibinfo{author}{\bibfnamefont{M.}~\bibnamefont{Baiesi}},
  \bibinfo{author}{\bibfnamefont{C.}~\bibnamefont{Maes}}, \bibnamefont{and}
  \bibinfo{author}{\bibfnamefont{B.}~\bibnamefont{Wynants}},
  \bibinfo{journal}{Phys. Rev. Lett.} \textbf{\bibinfo{volume}{103}},
  \bibinfo{pages}{010602} (\bibinfo{year}{2009}).

\bibitem[{\citenamefont{Chetrite and Gawedzki}(2009)}]{chet09}
\bibinfo{author}{\bibfnamefont{R.}~\bibnamefont{Chetrite}} \bibnamefont{and}
  \bibinfo{author}{\bibfnamefont{K.}~\bibnamefont{Gawedzki}},
  \bibinfo{journal}{J. Stat. Phys.} \textbf{\bibinfo{volume}{137}},
  \bibinfo{pages}{890} (\bibinfo{year}{2009}).

\bibitem[{\citenamefont{Prost et~al.}(2009)\citenamefont{Prost, Joanny, and
  Parrondo}}]{pros09}
\bibinfo{author}{\bibfnamefont{J.}~\bibnamefont{Prost}},
  \bibinfo{author}{\bibfnamefont{J.-F.} \bibnamefont{Joanny}},
  \bibnamefont{and} \bibinfo{author}{\bibfnamefont{J.~M.~R.}
  \bibnamefont{Parrondo}}, \bibinfo{journal}{Phys. Rev. Lett.}
  \textbf{\bibinfo{volume}{103}}, \bibinfo{eid}{090601} (\bibinfo{year}{2009}).

\bibitem[{\citenamefont{Speck}(2010)}]{spec10}
\bibinfo{author}{\bibfnamefont{T.}~\bibnamefont{Speck}},
  \bibinfo{journal}{Prog. Theor. Phys. Suppl.} \textbf{\bibinfo{volume}{184}},
  \bibinfo{pages}{248} (\bibinfo{year}{2010}).

\bibitem[{\citenamefont{Seifert}(2011)}]{seif11}
\bibinfo{author}{\bibfnamefont{U.}~\bibnamefont{Seifert}},
  \bibinfo{journal}{Eur. Phys. J. E} \textbf{\bibinfo{volume}{34}},
  \bibinfo{pages}{1} (\bibinfo{year}{2011}).

\bibitem[{\citenamefont{Warren and Allen}(2012)}]{warr12}
\bibinfo{author}{\bibfnamefont{P.~B.} \bibnamefont{Warren}} \bibnamefont{and}
  \bibinfo{author}{\bibfnamefont{R.~J.} \bibnamefont{Allen}},
  \bibinfo{journal}{Phys. Rev. Lett.} \textbf{\bibinfo{volume}{109}},
  \bibinfo{pages}{250601} (\bibinfo{year}{2012}).

\bibitem[{\citenamefont{Marconi et~al.}(2008)\citenamefont{Marconi, Puglisi,
  Rondoni, and Vulpiani}}]{marc08}
\bibinfo{author}{\bibfnamefont{U.~M.~B.} \bibnamefont{Marconi}},
  \bibinfo{author}{\bibfnamefont{A.}~\bibnamefont{Puglisi}},
  \bibinfo{author}{\bibfnamefont{L.}~\bibnamefont{Rondoni}}, \bibnamefont{and}
  \bibinfo{author}{\bibfnamefont{A.}~\bibnamefont{Vulpiani}},
  \bibinfo{journal}{Phys. Rep.} \textbf{\bibinfo{volume}{461}},
  \bibinfo{pages}{111} (\bibinfo{year}{2008}).

\bibitem[{\citenamefont{Seifert}(2012)}]{seif12}
\bibinfo{author}{\bibfnamefont{U.}~\bibnamefont{Seifert}},
  \bibinfo{journal}{Rep. Prog. Phys.} \textbf{\bibinfo{volume}{75}},
  \bibinfo{pages}{126001} (\bibinfo{year}{2012}).

\bibitem[{\citenamefont{Baiesi and Maes}(2013)}]{baie13}
\bibinfo{author}{\bibfnamefont{M.}~\bibnamefont{Baiesi}} \bibnamefont{and}
  \bibinfo{author}{\bibfnamefont{C.}~\bibnamefont{Maes}}, \bibinfo{journal}{New
  Journal of Physics} \textbf{\bibinfo{volume}{15}}, \bibinfo{pages}{013004}
  (\bibinfo{year}{2013}).

\bibitem[{\citenamefont{Chatelain}(2003)}]{chat03}
\bibinfo{author}{\bibfnamefont{C.}~\bibnamefont{Chatelain}},
  \bibinfo{journal}{J. Phys. A: Math. Gen.} \textbf{\bibinfo{volume}{36}},
  \bibinfo{pages}{10739} (\bibinfo{year}{2003}).

\bibitem[{\citenamefont{Diezemann}(2005)}]{diez05}
\bibinfo{author}{\bibfnamefont{G.}~\bibnamefont{Diezemann}},
  \bibinfo{journal}{Phys. Rev. E} \textbf{\bibinfo{volume}{72}},
  \bibinfo{pages}{011104} (\bibinfo{year}{2005}).

\bibitem[{\citenamefont{Berthier}(2007)}]{bert07}
\bibinfo{author}{\bibfnamefont{L.}~\bibnamefont{Berthier}},
  \bibinfo{journal}{Phys. Rev. Lett.} \textbf{\bibinfo{volume}{98}},
  \bibinfo{pages}{220601} (\bibinfo{year}{2007}).

\bibitem[{\citenamefont{Seifert and Speck}(2010)}]{seif10}
\bibinfo{author}{\bibfnamefont{U.}~\bibnamefont{Seifert}} \bibnamefont{and}
  \bibinfo{author}{\bibfnamefont{T.}~\bibnamefont{Speck}},
  \bibinfo{journal}{EPL} \textbf{\bibinfo{volume}{89}}, \bibinfo{pages}{10007}
  (\bibinfo{year}{2010}).

\bibitem[{\citenamefont{Lecomte et~al.}(2007)\citenamefont{Lecomte,
  Appert-Rolland, and van Wijland}}]{leco07a}
\bibinfo{author}{\bibfnamefont{V.}~\bibnamefont{Lecomte}},
  \bibinfo{author}{\bibfnamefont{C.}~\bibnamefont{Appert-Rolland}},
  \bibnamefont{and} \bibinfo{author}{\bibfnamefont{F.}~\bibnamefont{van
  Wijland}}, \bibinfo{journal}{J. Stat. Phys.} \textbf{\bibinfo{volume}{127}},
  \bibinfo{pages}{51} (\bibinfo{year}{2007}).

\bibitem[{\citenamefont{Garrahan et~al.}(2009)\citenamefont{Garrahan, Jack,
  Lecomte, Pitard, van Duijvendijk, and van Wijland}}]{garr09}
\bibinfo{author}{\bibfnamefont{J.~P.} \bibnamefont{Garrahan}},
  \bibinfo{author}{\bibfnamefont{R.~L.} \bibnamefont{Jack}},
  \bibinfo{author}{\bibfnamefont{V.}~\bibnamefont{Lecomte}},
  \bibinfo{author}{\bibfnamefont{E.}~\bibnamefont{Pitard}},
  \bibinfo{author}{\bibfnamefont{K.}~\bibnamefont{van Duijvendijk}},
  \bibnamefont{and} \bibinfo{author}{\bibfnamefont{F.}~\bibnamefont{van
  Wijland}}, \bibinfo{journal}{J. Phys. A: Math. Theor.}
  \textbf{\bibinfo{volume}{42}}, \bibinfo{pages}{075007}
  (\bibinfo{year}{2009}).

\bibitem[{\citenamefont{Turner et~al.}(2014)\citenamefont{Turner, Speck, and
  Garrahan}}]{turn14}
\bibinfo{author}{\bibfnamefont{R.~M.} \bibnamefont{Turner}},
  \bibinfo{author}{\bibfnamefont{T.}~\bibnamefont{Speck}}, \bibnamefont{and}
  \bibinfo{author}{\bibfnamefont{J.~P.} \bibnamefont{Garrahan}},
  \bibinfo{journal}{J. Stat. Mech.: Theor. Exp.} p. \bibinfo{pages}{P09017}
  (\bibinfo{year}{2014}).

\bibitem[{\citenamefont{Maes and Neto\v{c}n\'{y}}(2008)}]{maes08}
\bibinfo{author}{\bibfnamefont{C.}~\bibnamefont{Maes}} \bibnamefont{and}
  \bibinfo{author}{\bibfnamefont{K.}~\bibnamefont{Neto\v{c}n\'{y}}},
  \bibinfo{journal}{EPL} \textbf{\bibinfo{volume}{82}}, \bibinfo{pages}{30003}
  (\bibinfo{year}{2008}).

\bibitem[{\citenamefont{Merolle et~al.}(2005)\citenamefont{Merolle, Garrahan,
  and Chandler}}]{mero05}
\bibinfo{author}{\bibfnamefont{M.}~\bibnamefont{Merolle}},
  \bibinfo{author}{\bibfnamefont{J.~P.} \bibnamefont{Garrahan}},
  \bibnamefont{and} \bibinfo{author}{\bibfnamefont{D.}~\bibnamefont{Chandler}},
  \bibinfo{journal}{Proc. Natl. Acad. Sci. U.S.A.}
  \textbf{\bibinfo{volume}{102}}, \bibinfo{pages}{10837}
  (\bibinfo{year}{2005}).

\bibitem[{\citenamefont{Garrahan and Lesanovsky}(2010)}]{garr10}
\bibinfo{author}{\bibfnamefont{J.~P.} \bibnamefont{Garrahan}} \bibnamefont{and}
  \bibinfo{author}{\bibfnamefont{I.}~\bibnamefont{Lesanovsky}},
  \bibinfo{journal}{Phys. Rev. Lett.} \textbf{\bibinfo{volume}{104}},
  \bibinfo{pages}{160601} (\bibinfo{year}{2010}).

\bibitem[{\citenamefont{Chetrite and Touchette}(2013)}]{chet13}
\bibinfo{author}{\bibfnamefont{R.}~\bibnamefont{Chetrite}} \bibnamefont{and}
  \bibinfo{author}{\bibfnamefont{H.}~\bibnamefont{Touchette}},
  \bibinfo{journal}{Phys. Rev. Lett.} \textbf{\bibinfo{volume}{111}},
  \bibinfo{pages}{120601} (\bibinfo{year}{2013}).

\bibitem[{\citenamefont{Chetrite and Touchette}(2014)}]{chet14}
\bibinfo{author}{\bibfnamefont{R.}~\bibnamefont{Chetrite}} \bibnamefont{and}
  \bibinfo{author}{\bibfnamefont{H.}~\bibnamefont{Touchette}},
  \bibinfo{journal}{Ann. Henri Poincar\'e} \textbf{\bibinfo{volume}{16}},
  \bibinfo{pages}{2005} (\bibinfo{year}{2014}).

\bibitem[{\citenamefont{Szavits-Nossan and Evans}(2015)}]{szav15}
\bibinfo{author}{\bibfnamefont{J.}~\bibnamefont{Szavits-Nossan}}
  \bibnamefont{and} \bibinfo{author}{\bibfnamefont{M.~R.} \bibnamefont{Evans}},
  \bibinfo{journal}{J. Stat. Mech.} \textbf{\bibinfo{volume}{2015}},
  \bibinfo{pages}{P12008} (\bibinfo{year}{2015}).

\bibitem[{\citenamefont{Garrahan}(2016)}]{garr16}
\bibinfo{author}{\bibfnamefont{J.~P.} \bibnamefont{Garrahan}},
  \bibinfo{journal}{J. Stat. Mech.: Theor. Exp.}
  \textbf{\bibinfo{volume}{2016}}, \bibinfo{pages}{073208}
  (\bibinfo{year}{2016}).

\bibitem[{\citenamefont{Martyushev and Seleznev}(2006)}]{mart06}
\bibinfo{author}{\bibfnamefont{L.}~\bibnamefont{Martyushev}} \bibnamefont{and}
  \bibinfo{author}{\bibfnamefont{V.}~\bibnamefont{Seleznev}},
  \bibinfo{journal}{Phys. Rep.} \textbf{\bibinfo{volume}{426}},
  \bibinfo{pages}{1} (\bibinfo{year}{2006}).

\bibitem[{\citenamefont{Sasa and Tasaki}(2006)}]{sasa06}
\bibinfo{author}{\bibfnamefont{S.-i.} \bibnamefont{Sasa}} \bibnamefont{and}
  \bibinfo{author}{\bibfnamefont{H.}~\bibnamefont{Tasaki}},
  \bibinfo{journal}{J. Stat. Phys.} \textbf{\bibinfo{volume}{125}},
  \bibinfo{pages}{125} (\bibinfo{year}{2006}).

\bibitem[{\citenamefont{Jack et~al.}(2008)\citenamefont{Jack, Kelsey, Garrahan,
  and Chandler}}]{jack08}
\bibinfo{author}{\bibfnamefont{R.~L.} \bibnamefont{Jack}},
  \bibinfo{author}{\bibfnamefont{D.}~\bibnamefont{Kelsey}},
  \bibinfo{author}{\bibfnamefont{J.~P.} \bibnamefont{Garrahan}},
  \bibnamefont{and} \bibinfo{author}{\bibfnamefont{D.}~\bibnamefont{Chandler}},
  \bibinfo{journal}{Phys. Rev. E} \textbf{\bibinfo{volume}{78}},
  \bibinfo{eid}{011506} (\bibinfo{year}{2008}).

\bibitem[{\citenamefont{Onsager}(1931)}]{onsa31}
\bibinfo{author}{\bibfnamefont{L.}~\bibnamefont{Onsager}},
  \bibinfo{journal}{Phys. Rev.} \textbf{\bibinfo{volume}{37}},
  \bibinfo{pages}{405} (\bibinfo{year}{1931}).

\bibitem[{\citenamefont{Touchette}(2009)}]{touc09}
\bibinfo{author}{\bibfnamefont{H.}~\bibnamefont{Touchette}},
  \bibinfo{journal}{Phys. Rep.} \textbf{\bibinfo{volume}{478}},
  \bibinfo{pages}{1} (\bibinfo{year}{2009}).

\bibitem[{\citenamefont{Jack and Sollich}(2010)}]{jack10a}
\bibinfo{author}{\bibfnamefont{R.~L.} \bibnamefont{Jack}} \bibnamefont{and}
  \bibinfo{author}{\bibfnamefont{P.}~\bibnamefont{Sollich}},
  \bibinfo{journal}{Prog. Theor. Phys. Suppl.} \textbf{\bibinfo{volume}{184}},
  \bibinfo{pages}{304} (\bibinfo{year}{2010}).

\bibitem[{\citenamefont{Speck and Garrahan}(2011)}]{spec11}
\bibinfo{author}{\bibfnamefont{T.}~\bibnamefont{Speck}} \bibnamefont{and}
  \bibinfo{author}{\bibfnamefont{J.}~\bibnamefont{Garrahan}},
  \bibinfo{journal}{Eur. Phys. J. B} \textbf{\bibinfo{volume}{79}},
  \bibinfo{pages}{1} (\bibinfo{year}{2011}).

\bibitem[{\citenamefont{Nyawo and Touchette}(2016)}]{nyaw16}
\bibinfo{author}{\bibfnamefont{P.~T.} \bibnamefont{Nyawo}} \bibnamefont{and}
  \bibinfo{author}{\bibfnamefont{H.}~\bibnamefont{Touchette}},
  \bibinfo{journal}{arXiv:1606.02602}  (\bibinfo{year}{2016}).

\bibitem[{\citenamefont{Blickle et~al.}(2007)\citenamefont{Blickle, Speck,
  Lutz, Seifert, and Bechinger}}]{blic07}
\bibinfo{author}{\bibfnamefont{V.}~\bibnamefont{Blickle}},
  \bibinfo{author}{\bibfnamefont{T.}~\bibnamefont{Speck}},
  \bibinfo{author}{\bibfnamefont{C.}~\bibnamefont{Lutz}},
  \bibinfo{author}{\bibfnamefont{U.}~\bibnamefont{Seifert}}, \bibnamefont{and}
  \bibinfo{author}{\bibfnamefont{C.}~\bibnamefont{Bechinger}},
  \bibinfo{journal}{Phys. Rev. Lett.} \textbf{\bibinfo{volume}{98}},
  \bibinfo{pages}{210601} (\bibinfo{year}{2007}).

\bibitem[{\citenamefont{Gomez-Solano et~al.}(2009)\citenamefont{Gomez-Solano,
  Petrosyan, Ciliberto, Chetrite, and Gawedzki}}]{gome09}
\bibinfo{author}{\bibfnamefont{J.~R.} \bibnamefont{Gomez-Solano}},
  \bibinfo{author}{\bibfnamefont{A.}~\bibnamefont{Petrosyan}},
  \bibinfo{author}{\bibfnamefont{S.}~\bibnamefont{Ciliberto}},
  \bibinfo{author}{\bibfnamefont{R.}~\bibnamefont{Chetrite}}, \bibnamefont{and}
  \bibinfo{author}{\bibfnamefont{K.}~\bibnamefont{Gawedzki}},
  \bibinfo{journal}{Phys. Rev. Lett.} \textbf{\bibinfo{volume}{103}},
  \bibinfo{pages}{040601} (\bibinfo{year}{2009}).

\bibitem[{\citenamefont{Gomez-Solano et~al.}(2011)\citenamefont{Gomez-Solano,
  Petrosyan, Ciliberto, and Maes}}]{gome11}
\bibinfo{author}{\bibfnamefont{J.~R.} \bibnamefont{Gomez-Solano}},
  \bibinfo{author}{\bibfnamefont{A.}~\bibnamefont{Petrosyan}},
  \bibinfo{author}{\bibfnamefont{S.}~\bibnamefont{Ciliberto}},
  \bibnamefont{and} \bibinfo{author}{\bibfnamefont{C.}~\bibnamefont{Maes}},
  \bibinfo{journal}{J. Stat. Mech.: Theor. Exp.}
  \textbf{\bibinfo{volume}{2011}}, \bibinfo{pages}{P01008}
  (\bibinfo{year}{2011}).

\bibitem[{\citenamefont{Basu and Maes}(2015)}]{basu15}
\bibinfo{author}{\bibfnamefont{U.}~\bibnamefont{Basu}} \bibnamefont{and}
  \bibinfo{author}{\bibfnamefont{C.}~\bibnamefont{Maes}}, \bibinfo{journal}{J.
  Phys.: Conf. Ser.} \textbf{\bibinfo{volume}{638}}, \bibinfo{pages}{012001}
  (\bibinfo{year}{2015}).

\bibitem[{\citenamefont{Lebowitz and Spohn}(1999)}]{lebo99}
\bibinfo{author}{\bibfnamefont{J.~L.} \bibnamefont{Lebowitz}} \bibnamefont{and}
  \bibinfo{author}{\bibfnamefont{H.}~\bibnamefont{Spohn}}, \bibinfo{journal}{J.
  Stat. Phys.} \textbf{\bibinfo{volume}{95}}, \bibinfo{pages}{333}
  (\bibinfo{year}{1999}).

\bibitem[{\citenamefont{Mehl et~al.}(2008)\citenamefont{Mehl, Speck, and
  Seifert}}]{mehl08}
\bibinfo{author}{\bibfnamefont{J.}~\bibnamefont{Mehl}},
  \bibinfo{author}{\bibfnamefont{T.}~\bibnamefont{Speck}}, \bibnamefont{and}
  \bibinfo{author}{\bibfnamefont{U.}~\bibnamefont{Seifert}},
  \bibinfo{journal}{Phys. Rev. E} \textbf{\bibinfo{volume}{78}},
  \bibinfo{pages}{011123} (\bibinfo{year}{2008}).

\bibitem[{\citenamefont{Speck and Chandler}(2012)}]{spec12}
\bibinfo{author}{\bibfnamefont{T.}~\bibnamefont{Speck}} \bibnamefont{and}
  \bibinfo{author}{\bibfnamefont{D.}~\bibnamefont{Chandler}},
  \bibinfo{journal}{J. Chem. Phys.} \textbf{\bibinfo{volume}{136}},
  \bibinfo{pages}{184509} (\bibinfo{year}{2012}).

\bibitem[{\citenamefont{Kampen}(1981)}]{kampen}
\bibinfo{author}{\bibfnamefont{N.~G.~V.} \bibnamefont{Kampen}},
  \emph{\bibinfo{title}{Stochastic Processes in Physics and Chemistry}}
  (\bibinfo{publisher}{Elsevier}, \bibinfo{address}{Amsterdam},
  \bibinfo{year}{1981}).

\bibitem[{\citenamefont{Abramowitz and Stegun}(1972)}]{abramowitz}
\bibinfo{editor}{\bibfnamefont{M.}~\bibnamefont{Abramowitz}} \bibnamefont{and}
  \bibinfo{editor}{\bibfnamefont{I.~A.} \bibnamefont{Stegun}}, eds.,
  \emph{\bibinfo{title}{Handbook of Mathematical Functions}}
  (\bibinfo{publisher}{Dover}, \bibinfo{address}{New York},
  \bibinfo{year}{1972}), \bibinfo{edition}{9th} ed.

\end{thebibliography}
\end{document}